\documentclass[twocolumn,showpacs,aps,epsfig,nofootinbib]{revtex4}

\usepackage{graphicx}
\usepackage{epstopdf}
\usepackage{latexsym}
\usepackage{amssymb}
\usepackage{color}

\usepackage[center]{subfigure}

\begin{document}

 \newcommand{\bq}{\begin{equation}}
 \newcommand{\eq}{\end{equation}}
 \newcommand{\bqn}{\begin{eqnarray}}
 \newcommand{\eqn}{\end{eqnarray}}
 \newcommand{\nb}{\nonumber}
 \newcommand{\lb}{\label}
\newcommand{\PRL}{Phys. Rev. Lett.}
\newcommand{\PL}{Phys. Lett.}
\newcommand{\PR}{Phys. Rev.}
\newcommand{\CQG}{Class. Quantum Grav.}

\title{Gravitational collapse and formation of universal horizons}

\author{Miao Tian $^{a,b}$}

\author{Xinwen Wang $^{b}$}

\author{ M.F. da Silva $^{c}$}

\author{Anzhong Wang $^{b, c, d}$ \footnote{The corresponding author}}
\email{Anzhong_Wang@baylor.edu}

\affiliation{$^{a}$ School of Mathematics and Physics, Lanzhou Jiaotong University,
Lanzhou 730070, China\\
$^{b}$ GCAP-CASPER, Physics Department, Baylor
University, Waco, TX 76798-7316, USA\\
$^{c}$   Departamento de F\'isica Te\'orica, Universidade do Estado do Rio de Janeiro, Rua S\~ao Francisco Xavier 524, Maracan\~a,
CEP 20550Ð013, Rio de Janeiro, RJ, Brazil\\
$^{d}$ Institute  for Advanced Physics $\&$ Mathematics,
Zhejiang University of
Technology, Hangzhou 310032,  China
}

\date{\today}

\begin{abstract}

In this paper, we first generalize the definition of stationary universal horizons to dynamical ones, and then show that
(dynamical) universal horizons can be formed from realistic gravitational collapse. This is done   by constructing
analytical models of a collapsing spherically symmetric star with finite thickness in Einstein-aether theory.

\end{abstract}

\pacs{04.50.Kd, 04.70.Bw, 04.20.Jb, 97.60.-s} 

\maketitle

\section{ Introduction  }

\renewcommand{\theequation}{1.\arabic{equation}} \setcounter{equation}{0}

The invariance under the Lorentz symmetry group is a cornerstone of modern physics, and is strongly supported by observations.
In fact, all the experiments carried out so far are consistent with it  \cite{Liberati13}, and no evidence to show that such a symmetry
must be  broken at certain energy scales, although  it is arguable that such
constraints in  the gravitational sector are much weaker than those in the matter sector    \cite{LZbreaking}.

Nevertheless, there are various  reasons to construct gravitational theories  with broken Lorentz invariance (LI) \cite{LACW}.
In particular, our understanding of space-times at Plank scale is still highly limited, and the renomalizability and unitarity of gravity often
lead to the violation of LI \cite{QGs}. One concrete example is the Ho\v{r}ava theory of quantum gravity \cite{Horava},  in which the LI  is broken
via the anisotropic scaling between time and space  in the ultraviolet (UV),
\bq
\lb{0.1}
t \rightarrow b^{-z} t,\; \;\; x^{i} \rightarrow b^{-1} x^{i}, \; (i = 1, 2, 3),
\eq
where $z$ denotes the dynamical  critical exponent.  This is a reminiscent of Lifshitz scalars  in condensed matter physics \cite{Lifshitz}, hence the theory is often referred to as the
Ho\v{r}ava-Lifshitz (HL)  quantum gravity at a  Lifshitz  fixed point. The anisotropic scaling (\ref{0.1})  provides  a crucial mechanism: The gravitational
action can be constructed in such a way that only  higher-dimensional spatial (but not time) derivative operators are included, so that the UV behavior of the theory 
 is dramatically improved. In particular,  for $z \ge 3$   it  becomes  power-counting renormalizable \cite{Horava,Visser}. The exclusion of high-dimensional time derivative operators,
 on the other hand,   prevents the ghost instability, whereby the unitarity of the theory is assured \cite{Stelle}. In the infrared (IR) the lower dimensional operators
take over,  and  a healthy low-energy limit is presumably resulted \footnote{It should be emphasized  that, the breaking of LI can have significant effects on the low-energy physics  through
the interactions between gravity and matter, no matter how high the scale of symmetry breaking is
\cite{Collin04}. Recently, Pospelov and Tamarit proposed a mechanism of SUSY breaking by coupling a Lorentz-invariant supersymmetric matter sector to non-supersymmetric gravitational
interactions with Lifshitz scaling, and showed that it can lead to a consistent HL gravity \cite{PT14}.}.
It is remarkable to note that, despite of the stringent observational  constraints of the violation of the LI \cite{Liberati13}, the nonrelativistic  general covariant  HL gravity constructed 
in \cite{ZWWS} is consistent with all the solar system tests \cite{Will,LMWZ} and
cosmology  \cite{cosmo,ZHW}.  In addition,  it  has been recently embedded   in string theory via the
 nonrelativistic AdS/CFT correspondence  \cite{JK}. Another version of the HL gravity,  the health extension \cite{BPS}, is also self-consistent and passes all the solar system, astrophysical 
 and cosmological  tests \footnote{ In fact, in the IR the theory can be identified with the hypersurface-orthogonal Einstein-aether theory \cite{EA} in a particular gauge \cite{Jacob,Wang13,LACW}, whereby the 
 consistence of the theory with observations  can be deduced.}.

Another example that violates LI  is the Einstein-aether theory,  in which  the breaking is realized by a timelike vector field, while the gravitational action is still generally covariant \cite{EA}.
This theory is consistent with all the solar system tests \cite{EA} and binary pulsar observations \cite{Yagi}.

However, once the LI is broken, speeds of particles can be greater than that of light. In particular,  the dispersion relation generically becomes nonlinear \cite{ZHW}, 
\bq
\lb{0.2}
E^2 = c_{p}^2 p^2\left(1 + \alpha_1 \left(\frac{p}{M_{*}}\right)^2 +  \alpha_2  \left(\frac{p}{M_{*}}\right)^4\right),
\eq
where $E$ and $p$ are the energy and momentum of the particle considered, and $c_p, \; \alpha_i$ are coefficients, depending on the
species of the particle, while $M_{*}$ denotes the suppression energy  scale of the higher-dimensional operators. Then,   one can see that
both phase and group velocities of the particles are unbounded  with the increase of energy.   This suggests that black holes may not exist at all in theories with broken LI, and makes
such theories questionable, as observations strongly indicate that black holes exist in our universe \cite{NM}.

Lately, a potential breakthrough  was the discovery that   there still exist absolute causal boundaries, the
so-called {\em universal horizons},  in the   theories  with broken LI  \cite{BS11}.   Particles even with infinitely large velocities    would just move around on these
boundaries and  cannot escape to infinity \cite{UHs}.  The universal horizon radiates like a blackbody at a fixed temperature,
and obeys  the first law of black hole mechanics  \cite{BBM}.

 Recently,  we studied the existence of universal horizons in  the three well-known black hole solutions, the Schwarzschild, Schwarzschild anti-de Sitter, and Reissner-Nordstr\"om,
 and found that   in all  of them  universal horizons always exist  inside their Killing horizons \cite{LGSW}.  In
particular,   the  peeling-off behavior of the globally timelike khronon field  $u_{\mu}$ was found only at the universal horizons, whereby
the surface gravity $\kappa_{\text{peeling}}$ is calculated and  found  equal to \cite{CLMV},
\bq
\lb{0.3}
 \kappa_{\text{UH}} \equiv  \frac{1}{2} u^{\alpha} D_{\alpha} \left(u_{\lambda}  \zeta^{\lambda}\right),
\eq
 where $\zeta^{\mu}$ and $D_{\mu}$ denote, respectively, the  time translation Killing field and covariant derivative with respect to the given space-time metric $g_{\mu\nu}\;
(\mu, \nu = 0, 1, 2, 3)$. For the Schwarzschild solution, the universal horizon and surface gravity are given, respectively,  by \cite{LGSW},
\bq
 \lb{0.4}
 R_{UH}^{\text{Sch.}} = \frac{3r_s}{4},\;\;\;  \kappa_{\text{peeling}}^{\text{Sch.}} = \left(\frac{2}{3}\right)^{3/2}\frac{1}{r_s},
 \eq
 where $r_s$ denotes the Schwarzschild radius.

 In this paper, we shall study the formation of the universal horizons from realistic gravitational collapse of a spherically symmetric star
  \footnote{Here ``realistic" means that  the collapsing object  satisfies at least  the weak
 energy condition \cite{HE73}.}. To be more concrete, we shall consider such a collapsing object in the Einstein-aether theory \cite{EA}. To make the problem tractable, we further
 assume that the effects of the aether are negligible, so the space-time outside of the star is still  described by the Schwarzschild solution, while inside the star we assume that the
 distribution of the matter is homogeneous and isotropic, so the internal space-time is that of  the Friedmann-Robertson-Walker (FRW). Although the model is very ideal, it is sufficient to
 serve our current purposes, that is, to show explicitly that universal horizons can be  formed from realistic gravitational collapse.

 Specifically, the paper is organized as follows: In Sec. II,
 we shall present a brief review on the definition of the stationary universal horizons, and then
 generalize it to dynamical spacetimes. This is realized by replacing Killing horizons by apparent horizons \cite{Hayd,Wang}, and in the stationary limit, the latter reduces to the former.
 In Sec. III, we study a collapsing spherically symmetric star with a finite thickness in the framework of the Einstein-aether theory. When the effects of the aether are negligible, the
 vacuum space-time outside the star  is uniquely described by the Schwarzschild solution and the junction condition across the surface of the star reduces to those of Israel \cite{IC66}.
 Once this is done, we find that the khronon equation can be solved analytically when the speed of the khronon is infinitely large, for which the sound horizon of the khronon coincides with
 the universal horizon. It is remarkable that this is also the case for the
 Schwarzschild solution \cite{LGSW}, for  which the universal horizon and surface gravity are given   by  Eq.(\ref{0.4}). The paper is ended in Sec. IV, in which  our main conclusions are
 presented.

\section{Dynamical  universal horizons and black holes}

\renewcommand{\theequation}{2.\arabic{equation}} \setcounter{equation}{0}

A necessary condition for the existence of  a universal horizon of a given space-time is the existence of  a globally time-like foliation \cite{BS11,LACW,LGSW}. This
foliation is usually characterized by a scalar field $\phi$, dubbed
khronon  \cite{BS11}, and the  normal vector $u_{\mu}$ of the foliation   is always time-like,
\bq
\lb{1.2}
u^{\lambda}u_{\lambda} = -1,
\eq
where
\bq
\lb{1.1}
u_{\mu} = \frac{\phi_{,\mu}}{\sqrt{X}}, \;\;\;  \phi_{,\mu} \equiv \frac{\partial \phi}{\partial x^{\mu}}, \;\;\;
 X \equiv -g^{\alpha\beta}\partial_{\alpha} \phi \partial_{\beta} \phi.
\eq
In this paper,  we choose the signature of the metric as ($-1, 1, 1, 1$).
It is important to note that such a defined  khronon field is unique only up to the following    gauge transformation,
\bq
\lb{1.3}
\tilde{\phi}  =  {\cal{F}}(\phi),
\eq
where  ${\cal{F}}(\phi)$ is a monotonically increasing (or decreasing) and otherwise arbitrary function of $\phi$. Clearly, under the above gauge transformations, we have
$\tilde{u}_{\mu} = u_{\mu}$.

The khronon field $\phi$ is described by the action  \cite{EA},
\bqn
\lb{1.4}
S_{\phi} &=&  \int d^{4}x \sqrt{|g|}\Big[V_0\left(D_{\mu}u_{\nu}\right)^2 + V_0 \left(D_{\mu}u^{\mu}\right)^2\nb\\
&& ~~  + c_3   \left(D^{\mu}u^{\nu}\right)\left( D_{\nu}u_{\mu}\right)   - c_4 a^{\mu}a_{\mu} \Big],
\eqn
where  $c_i$'s are arbitrary constants, and
$a_{\mu} \equiv u^{\alpha}D_{\alpha}u_{\mu}$. The operator $D_{\mu}$ denotes the covariant derivative with respect to the background metric $g_{\mu\nu}$, as mentioned above.
Note that the above action is the most general one in the sense that the
resulting  differential equations in terms of   $u_{\mu}$ are  second-order \cite{EA}. However,
when $u_{\mu}$ is  written in the form  of Eq.(\ref{1.1}),   the relation
\bq
\lb{1.6}
u_{[\nu}D_{\alpha}u_{\beta]} = 0,
\eq
is  identically satisfied.  Then, it can be shown that only three of the four coupling constants $c_i$  are independent. In fact, from Eq.(\ref{1.6})  we find   \cite{EA},
\bq
\lb{1.7}
\Delta{\cal{L}}_{\phi} \equiv  a^{\mu}a_{\mu} + \big(D_{\alpha}u_{\beta}\big)\big(D^{\alpha}u^{\beta}\big) -   \big(D_{\alpha}u_{\beta}\big)\big(D^{\beta}u^{\alpha}\big) = 0.
\eq
Then, one can always add the term,
\bq
\lb{1.8}
\Delta{S}_{\phi} = c_0 \int{\sqrt{|g|} \; d^{4}x \Delta{\cal{L}}_{\phi}},
\eq
into $S_{\phi}$, where $c_0$ is an arbitrary constant. This is effectively to shift the coupling constants $c_i$ to ${c}_i'$, where
\bq
\lb{1.9}
{c}_{1}' = c_1 + c_0,\; {c}_{2}' = c_2,\; {c}_{3}' = c_3  -  c_0,\;
{c}_{4}' = c_4  - c_0.
\eq
Thus, by properly choosing $c_0$, one can always set one of $c_{i}\; (i = 1, 3, 4)$ to zero. However, in the following we shall leave this possibility open.

The variation of  $S_{\phi}$ with respect to $\phi$ yields
the khronon equation,
\bqn
\lb{1.11}
D_{\mu} {\cal{A}}^{\mu}  = 0,
\eqn
where \cite{Wang13},
\bqn
\lb{1.12}
{\cal{A}}^{\mu} &\equiv& \frac{\left(\delta^{\mu}_{\nu}  + u^{\mu}u_{\nu}\right)}{\sqrt{X}}\AE^{\nu},\nb\\
\AE^{\nu} &\equiv& D_{\gamma} J^{\gamma\nu} + c_4 a_{\gamma} D^{\nu}u^{\gamma},\nb\\
J^{\alpha}_{\;\;\;\mu} &\equiv&  \big(V_0g^{\alpha\beta}g_{\mu\nu} + V_0 \delta^{\alpha}_{\mu}\delta^{\beta}_{\nu}
+  c_3 \delta^{\alpha}_{\nu}\delta^{\beta}_{\mu}\nb\\
&&  ~~~ - c_4 u^{\alpha}u^{\beta} g_{\mu\nu}\big)D_{\beta}u^{\nu}.
\eqn

\subsection{ Universal Horizons in Stationary and Asymptotically flat Spacetimes}

In stationary and asymptotically flat spacetimes, there always exists a time translation Killing vector, $\zeta^{\mu}$, which is timelike asymptotically,
\bq
\lb{1.13a}
\zeta^{\lambda}\zeta_{\lambda} < 0,
\eq
for $r \rightarrow \infty$.
A {\em Killing horizon} is defined as the existence of a hypersurface on which the time translation Killing vector $\zeta^{\mu}$ becomes null,
\bq
\lb{1.10}
\zeta^{\lambda} \zeta_{\lambda} = 0.
\eq
On the other hand, a {\em  universal horizon} is defined as the existence of a hypersurface on which $\zeta^{\mu}$ becomes orthogonal to
$u_{\mu}$,
\bq
\lb{1.16}
u_{\lambda} \zeta^{\lambda} = 0.
\eq
Since $u_{\mu}$ is timelike globally, Eq.(\ref{1.16}) is possible only when $\zeta_{\mu}$ becomes spacelike. This can happen  only  inside the
apparent horizons, because only in that region $\zeta_{\mu}$ becomes  spacelike.

\subsection{Universal Horizons in Non-Stationary   Spacetimes}

To study the formation of universal horizons from gravitational collapse, we need first to generalize the above definition of the universal horizons to non-stationary spacetimes.
For the sake of simplicity, in the rest of this paper we shall restrict ourselves only to spherical space-times, and its generalization to other spacetimes is straightforward.

The metric for a specially symmetric space-time can  be cost in the form,
\bq
\lb{2.1}
ds^2 = g_{ij}dx^idx^j + {R}^2\left(x^i\right)d\Omega^2, \; (i, j = 0, 1),
\eq
in the spherical coordinates, $x^{\mu} = \left(x^0, x^1, \theta, \varpi\right), \; (\mu = 0, 1, 2, 3)$, where $d\Omega^2 = d\theta^2 + \sin^2\theta d\varpi^2$.

 The normal vector $n_{\mu}$ to the hypersurface $ {R} = C_0$ is given by,
\bq
\lb{2.3}
n_{\mu} \equiv \frac{\partial({R} -  C_0)}{\partial x^{\mu}} = \delta^{0}_{\mu} {R}_{0} + \delta^{1}_{\mu} {R}_{1},
\eq
where  $C_0$ is a constant and  ${R}_{i} \equiv \partial{R} /\partial x^i$. Setting
\bq
\lb{2.4}
\zeta^{\mu} = \delta^{0}_{\mu} {R}_{1} - \delta^{1}_{\mu} {R}_{0},
\eq
we can see that $\zeta^{\mu}$ is always orthogonal to $n_{\mu}$,
\bq
\lb{2.4a}
\zeta^{\lambda} n_{\lambda} = 0.
\eq

For spacetimes that are asymptotically flat there always exists a region, say, ${R} > {R}_{\infty}$,
in which $n_{\mu}$ and $\zeta^{\mu} $ are, respectively, space- and time-like, that is,
$\left. n_{\mu} n^{\mu}\right|_{{R} > {R}_{\infty}} > 0$ and $\left. \zeta^{\mu} \zeta_{\mu}\right|_{{R} > {R}_{\infty}} < 0$.
An apparent horizon may form at ${R}_{AH}$, at which
$n_{\mu} $ becomes null,
\bq
\lb{2.4b}
\left. n_{\lambda} n^{\lambda} \right|_{{R} = {R}_{AH}} = 0,
\eq
where $ {R}_{AH} <  {R}_{\infty} $. Then, in the internal region $ {R} < {R}_{AH}$,  $n_{\mu} $ becomes timelike.
Therefore, we have
\bq
\lb{2.4c}
n_{\lambda} n^{\lambda}  = \cases{ > 0, & ${R} > {R}_{AH}$, \cr
= 0, & ${R} = {R}_{AH}$, \cr
< 0, & ${R} < {R}_{AH}$. \cr}
\eq
Since Eq.(\ref{2.4a}) always holds, we must have
\bq
\lb{2.4d}
\zeta_{\lambda} \zeta^{\lambda}  = \cases{ < 0, & ${R} > {R}_{AH}$, \cr
= 0, & ${R} = {R}_{AH}$, \cr
> 0, & ${R} < {R}_{AH}$, \cr}
\eq
that is, $\zeta^{\mu}$  becomes null on the apparent horizon, and spacelike inside it.

We define a {\em dynamical universal horizon} as the hypersurface at which
\bq
\lb{2.16}
\left. u_{\lambda} \zeta^{\lambda}\right|_{{R} = {R}_{UH}} = 0.
\eq
Since $u_{\mu}$ is  globally timelike, Eq.(\ref{2.16}) is possible only when $\zeta_{\mu}$ is spacelike. Clearly, this is possible only  inside the
apparent horizons, that is, $  {R}_{UH} <  {R}_{AH}$.

In the static case, the apparent horizons defined above reduce to the Killing horizons, and the dynamical universal horizons defined by Eq.(\ref{2.16})
are identical to those given by Eq.(\ref{1.16}).

\section{Gravitational Collapse and Formation of Universal Horizons}
\renewcommand{\theequation}{3.\arabic{equation}} \setcounter{equation}{0}

Let us consider a collapsing star with a finite radius ${R}_{\Sigma}(\tau)$, where $\tau$ denotes the proper time of the surface of the star.
To our current purpose, we simply assume that the space-time inside the star is described by
  the  FRW  flat  metric,
 \bq
\lb{4.1}
 ds^2_{-} =-dt^2+a^2(t)\left(dr^2+ r^2d\Omega^2\right),\; ({R} \le {R}_{\Sigma}(\tau)),
 \eq
where ${R} =a(t)r$ is the geometric radius inside the collapsing star.
From Eq.(\ref{2.3}), the normal vector $n_{\mu}$ to the hypersurface $ {R} = C_0$  takes the form,
\bq
n_{\mu} = \dot{a}(t)r \delta^{0}_{\mu}+ a(t)\delta^{1}_{\mu},
\eq
where $\dot{a} \equiv da/dt$.   Then,  the corresponding vector $\zeta^{\mu}$ reads
\bq
\lb{4.3}
\zeta^{\mu} =a(t) \delta^{0}_{\mu} -\dot{a}(t)r  \delta^{1}_{\mu}.
\eq
According to Eq.(\ref{2.4b}) the apparent horizon locates at,
\bq
\lb{4.4}
r_{AH} = - \frac{1}{\dot{a}(t)}.
\eq
Note that in the collapsing case we have $\dot{a} < 0$.

The spacetime outside the collapsing star is vacuum. In the Einstein-aether theory \cite{EA}, if we consider the case where the effects of the aether field is negligible,
then the vacuum space-time will be that of the Schwarzschild, and in the  ingoing Eddington-Finkelstein coordinates $(v, {R}, \theta,\varpi)$ the metric takes the form,
\bq
\lb{4.13}
ds^2_{+}=-\left(1-\frac{2M}{{R}}\right)dv^2+2dv d{R}+ {R}^2d\Omega^2,\;\; ({R} \ge {R}_{\Sigma}(\tau)),
\eq
where $M$ denotes the total mass of the collapsing system, including that of the star surface, and has the dimension of length $L$, that is, $[M] = L$. The surface  $\Sigma$ of the star
can  be parameterized as,
\bqn
\lb{4.14}
r-r_\Sigma&=&0 \; (in\; V^-),\nb\\
{R}-{R}_\Sigma(v)&=&0 \; (in\; V^+),
\eqn
where $r_\Sigma$ is a constant, when we choose the internal coordinates $(t, r, \theta, \varpi)$ are comoving with the fluid of the collapsing star.
On the surface of the collapsing star,  the interior and exterior metrics reduce to
\bqn
\lb{4.15}
ds_{-} ^2|_{r=r_\Sigma}&=&-dt^2+a^2(t)r_\Sigma^2d\Omega^2\nb\\
&=& ds_{+} ^2|_{{R}= {R}_\Sigma(v)}\nb\\
&=& -\left(1-\frac{2M}{R_\Sigma(v)}-2\frac{dR_\Sigma(v)}{dv}\right)dv^2 \nb\\
&& + R_\Sigma(v)^2d\Omega^2\nb\\
&\equiv &-d\tau^2+R_\Sigma(\tau)^2d\Omega^2,
\eqn
where $R_\Sigma(\tau)$ is the geometric radius of the collapsing star, $v$ is a function of $\tau$, where $\tau$ denotes the proper time of the observers that are comoving with the collapsing surface
of the star.
In the current case, we have  $\tau=t$. Then,  we find
\bqn
\lb{4.16}
R_\Sigma(\tau)=a(\tau)r_\Sigma,\nb\\
\left(1-\frac{2M}{R_\Sigma}\right)\dot{v}^2-2\dot{R}_\Sigma\dot{v}-1=0,
\eqn
where a dot denotes the derivative with respect to $\tau$. However, since in the present case  we have $\tau = t$, so we shall not distinguish it from that with respect to $t$, used  above.
The extrinsic curvature tensor on the two sides of the surface defined by,
\bqn
\lb{4.17}
K^\pm_{ab}=-n^\pm_\alpha\Big[\frac{\partial^{2}x^\alpha_\pm}{\partial\xi^a\partial\xi^b}+{\Gamma^\pm_{\beta \delta}}^\alpha\frac{\partial
x^\beta_\pm}{\partial\xi^a}\frac{\partial x^\delta_\pm}{\partial\xi^b}\Big],
\eqn
 has the following non-vanishing components \cite{CS97} \footnote{Note the sign difference of the first term of $K^+_{\tau\tau}$ between the one obtained here and that obtained in
 \cite{CS97},  because of the sign difference of the cross term $dvdR$ in the external metric (\ref{4.13}).}
\bqn
\lb{4.18}
K^+_{\tau\tau}&=&-\frac{\ddot{v}}{\dot{v}}-\frac{M\dot{v}}{{R_\Sigma}^2},\nb\\
K^-_{\theta\theta}&=&sin^{-2}\theta K^-_{\varphi\varphi}=a(\tau)r_\Sigma,\\
K^+_{\theta\theta}&=&sin^{-2}\theta K^+_{\varphi\varphi}=(R_\Sigma-2M)\dot{v}-R_\Sigma\dot{R_\Sigma},\nb
\eqn
where
$n^\pm_\alpha$ are the normal vectors defined in the two faces of the surface (\ref{4.14}),
\bqn
\lb{4.18a}
n^{-}_\alpha &\equiv& \frac{\partial(r - r_{\Sigma})}{\partial x_{-}^{\alpha}} = \delta^{r}_{\alpha},\nb\\
n^{+}_\alpha &\equiv& \frac{\partial(R - R_{\Sigma}(v))}{\partial x_{+}^{\alpha}} = \delta^{R}_{\alpha} -\frac{dR_{\Sigma}(v)}{dv} \delta^{v}_{\alpha}.
\eqn
From the Israel junction  conditions  \cite{IC66},
\bqn
\lb{4.19}
\left[K_{ab}\right]^{-}-g_{ab}\left[K\right]^-=-8\pi\tau_{ab},
\eqn
we can get the surface energy-momentum tensor $\tau_{ab}$, where $\left[K_{ab}\right]^{-} \equiv K^+_{ab}-K^-_{ab},    \; [K]^- \equiv g^{ab}[K_{ab}]^-$, and $g_{ab}$ can be read off from Eq.(\ref{4.15}),
where $a, b = \tau, \theta, \varpi$.
Inserting  Eq.(\ref{4.18}) into the above equation,  we find that $\tau_{ab}$ can be written in the form
\bqn
\lb{4.20}
\tau_{ab}=\sigma w_a w_b +\eta(\theta_a \theta_b+\varpi_a \varpi_b),
\eqn
where $w_a,\theta_a$  and $ \varpi_a$ are unit vectors defined on the surface of the star, given, respectively,  by 
$w_a=\delta^\tau_a, \; \theta_a=R_\Sigma \delta^\theta_a, \varpi_a= R_\Sigma \sin\theta \delta^\varpi_a$,
and
\bqn
\lb{4.21}
\sigma&=&\frac{1}{4\pi r_\Sigma a}+\frac{\dot{a}}{4\pi a}+\frac{M\dot{v}}{2\pi r^2_\Sigma a^2}-\frac{\dot{v}}{4\pi r_\Sigma a},\\
\eta&=& \frac{\dot{v}}{8\pi r_\Sigma a}+\frac{\ddot{v}}{8\pi \dot{v}}-\frac{1}{8\pi r_\Sigma a}-\frac{\dot{a}}{8\pi a}-\frac{M\dot{v}}{8\pi r^2_\Sigma a^2},\nb
\eqn
here $\sigma$ is the surface energy density of the collapsing star, and $\eta$ its tangential pressure. Physically, they are often required to satisfy  certain energy conditions,
such as weak, strong and dominant \cite{HE73}, although in cosmology none of them seems necessarily to be satisfied \cite{Cosmo}.

On the other hand, inside the collapsing star, the khronon can be parametrized as,
\bqn
\lb{4.6}
u^\mu&=&\sqrt{1+a^2V^2} \delta^{\mu}_{0}+ V\delta^{\mu}_{1},\nb\\
u_\mu&=&-\sqrt{1+a^2V^2} \delta^{\mu}_{0}+ a^2V\delta^{\mu}_{1},
\eqn
where $V = V(t, r)$ is determined by the khronon equation (\ref{1.11}), which now  reduces to
\bq
\lb{4.5}
{\cal{A}}^{r}_{,r} + \frac{2{\cal{A}}^r}{r}+ {\cal{A}}^{t}_{,t} + \frac{3\dot{a}(t){\cal{A}}^t}{a(t)}= 0,
\eq
where
\bqn
 \lb{4.7}
{\cal{A}}^t&=&c_{123}{\cal{A}}^t_1+c_{14}{\cal{A}}^t_2+c_{13}{\cal{A}}^t_3\nb\\
{\cal{A}}^{r} &=& \frac{\sqrt{1+a^2V^2}} {a^2V} {\cal{A}}^t,\;\;\; {\cal{A}}^{\theta} = {\cal{A}}^{\phi} = 0,
\eqn
with $c_{ab}\equiv c_a +c_b,\; c_{abc}\equiv c_a +c_b + c_c$, and
 \bqn
\lb{4.7a}
{\cal{A}}^t_1&=&\frac{V}{r^2(1+a^2V^2)}(2a^4V^5 ( -1+2r^2a\ddot{a})\nb \\
&&+r(r V''+ V'(2+r a^2\sqrt{1+a^2V^2}\dot{V}))\nb\\
&&+V(-2-3r^2\dot{a}^2+3r^2a\ddot{a}+5r^2a\sqrt{1+a^2V^2}\dot{a}V'\nb\\
&&+r^2a^4\dot{V}^2+2ra^2\sqrt{1+a^2V^2}(\dot{V}+r\dot{V'}))\nb\\
&&+a^2V^3(-4+r(-2r\dot{a}^2+4ra\sqrt{1+a^2V^2}\dot{a}V'\nb\\
&&+a(7r\ddot{a}+2a\sqrt{1+a^2V^2}(\dot{V}+r\dot{V'}))))\nb\\
&&+ra^4V^4(2V'+r(V''+a(5\dot{a}\dot{V}+a\ddot{V})))\nb\\
&&+ra^2V^2(4V'+r(2V''+a(7\dot{a}\dot{V}+a\ddot{V}))) ),
\eqn
\bqn
\lb{4.7b}
{\cal{A}}^t_2&=&-\frac{V}{r(1+a^2V^2)}(4rV\dot{a}^2+ra^5V^4(2V\ddot{a}+5\dot{a}\dot{V}) \nb\\
&&+2a^3V^2(2V^2\sqrt{1+a^2V^2}\dot{a}+2rV(\ddot{a} \nb\\
&&+\sqrt{1+a^2V^2}\dot{a}V')+5r\dot{a}\dot{V})+a(4V^2\sqrt{1+a^2V^2}\dot{a}\nb\\
&&+ r V(2\ddot{a}+5\sqrt{1+a^2V^2}\dot{a}V')+5r\dot{a}\dot{V})\nb\\
&&+ra^6V^4\ddot{V}+a^4V^2(4rV^3\dot{a}^2+V^2(2V'+r V'')\nb\\
&&+2V\sqrt{1+a^2V^2}(\dot{V}+r\dot{V'})+2r\ddot{V})\nb\\
&&+a^2(8rV^3\dot{a}^2+V^2(2V'+r V'')+V(rV^2\nb\\
&&+2\sqrt{1+a^2V^2}(\dot{V}+r\dot{V'}))\nb\\
&&+r(\sqrt{1+a^2V^2}V'\dot{V}+\ddot{V})) ),\\
\nb\\
{\cal{A}}^t_3&=&2V^2(1+a^2V^2)(\dot{a}^2-a\ddot{a}).
\eqn
Here a prime denotes the derivative with respect to $r$. It is found  very difficult to solve Eq.(\ref{4.4}) for any given coupling constants $c_i$. However, when $c_{14}=0$, we obtain a particular solution,
\bqn
\lb{4.11}
V(t,r)=\frac{V_0 r}{a(t)}, \;\;\;
a(t)=a_0e^{-Ht},
\eqn
where $V_0, \; H$ and $a_0$ are  integration constants with $H> 0, \; a_0 > 0$.   It is remarkable to note that $c_{14} = 0$ corresponds to the case in which the speed of the
khronon becomes infinitely large $c_{\phi}^2 = c_{123}/c_{14} \rightarrow \infty$, a case that was also studied in \cite{BS11,LACW,LGSW}.

From the definition of the dynamical universal horizon Eq.(\ref{2.16}) and considering Eqs.(\ref{4.3}), (\ref{4.6}) and (\ref{4.11}), we find that the collapse always forms a universal horizons
inside the collapsing star, and its location  is given by,
\bq
\lb{4.12}
r_{UH}(t) = \sqrt{\frac{2}{\sqrt{V_0^4+4V_0^2 a_0^2 H^2 e^{-2H t}}  -V_0^2}}.
\eq

From Eq.(\ref{4.16}), on the other hand, we find that,
\bqn
\lb{4.22}
\dot{v}=\frac{-r^2_\Sigma a \dot{a}+\sqrt{r^2_\Sigma a^2+r^2_\Sigma a^2 \dot{a}^2-2a M r_\Sigma}}{2M-a r_\Sigma}.
\eqn
Substituting it together with $a(\tau)=a_0e^{-H\tau}$ back into Eqs.(\ref{4.21}), we obtain 
\bqn
\lb{4.23}
&\sigma&=\frac{1}{4\pi R_\Sigma}\left(1+{\cal{G}}\right),\nb\\
&\eta&= \frac{1}{8\pi R_\Sigma}\left(\frac{M-R_\Sigma -2H^2 R^3_\Sigma}{R_\Sigma {\cal{G}}} - 1\right), ~~~~~
\eqn
  where
\bq
\lb{4.23a}
R_\Sigma = a_0 r_\Sigma  e^{-H\tau},\;\;\;
{\cal{G}} \equiv \sqrt{1+H^2 R^2_\Sigma -\frac{2M}{R_\Sigma }}.
\eq
Obviously, to have both $\sigma$ and $\eta$ real, we must assume that $R_\Sigma \ge R_\Sigma^{Min}$, where $ R_\Sigma^{Min}$ is a root of the equation,
\bq
\lb{4.23b}
1+H^2 R^2_\Sigma -\frac{2M}{R_\Sigma } =  0.
\eq
When the star collapses to the point  $R_\Sigma\left(\tau_{Min}\right) =  R_\Sigma^{Min}$, the tangential pressure diverges, whereby a space-time singularity (with a finite radius)  is developed.
This represents the end of the collapse, as the space-time beyond this moment is not extendable.

\begin{figure}[tbp]
\centering
\includegraphics[width=8cm]{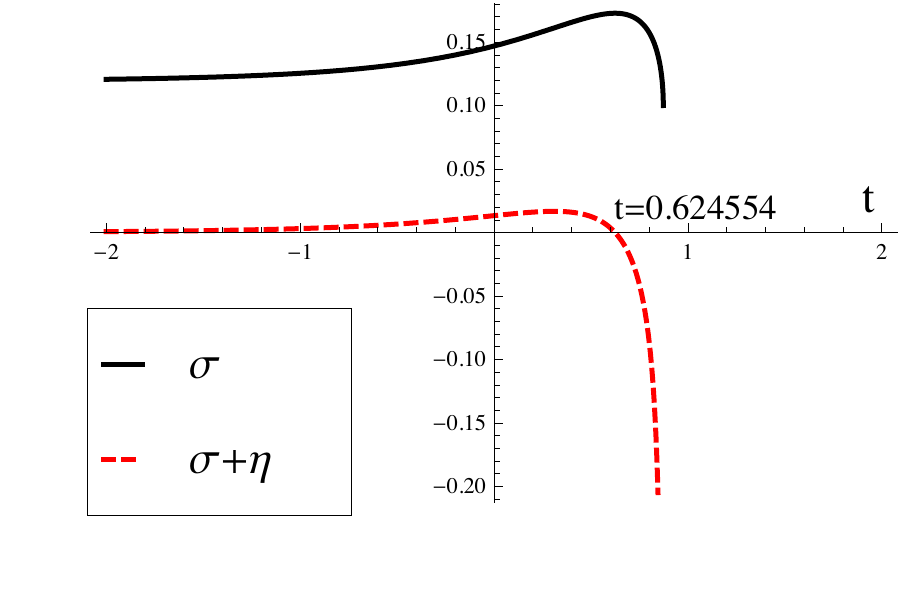}
\caption{The solid (black)  line  represents the surface energy density $\sigma$, while the dashed (red) line denotes the quantity $\sigma + \eta$.
When plotting these curves, we set $r_\Sigma=1, M = 1, a_0 = 3, V_0 = 0.6, H = 1.5$. At the moment  $t=0.624554$, we have
$\sigma+\eta=0$, $R_\Sigma=1.1756$, and $ R_{UH}= 1.24387$.}
 \label{fig1}
\end{figure}

\begin{figure}[tbp]
\centering
\includegraphics[width=8cm]{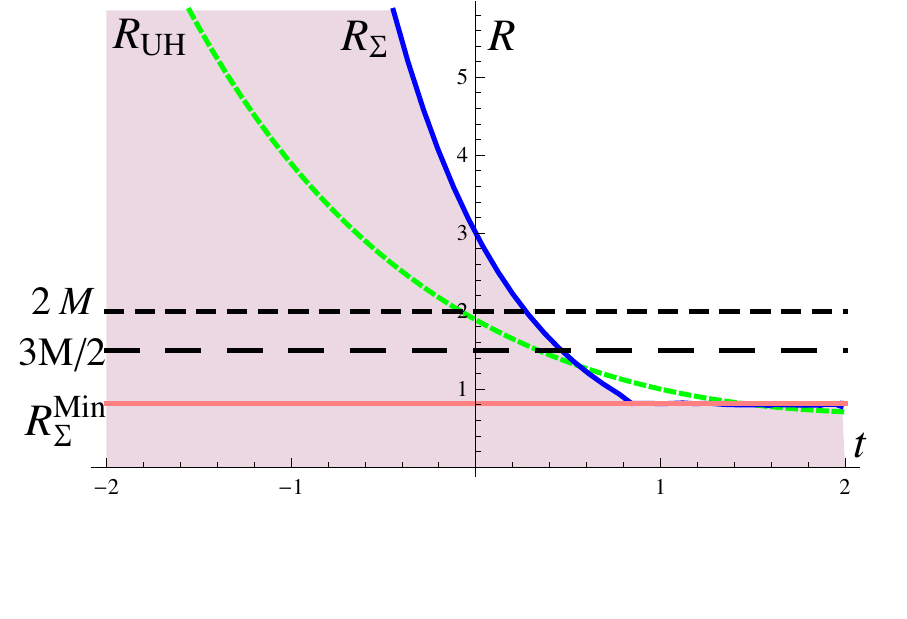}
\caption{Universal horizon $R_{UH}$, the surface  radius $R_{\Sigma}$ of the collapsing star vs $t$  for  $r_\Sigma=1, M = 1, a_0 = 3, 
V_0 = 0.6, H = 1.5$.  The horizontal lines $ R = 2M$ and $R = 3M/2$ are, respectively,  the Killing and universal horizons of the Schwarzschild vacuum solution. 
The collapse ends at the moment when $R_{\Sigma} = R^{Min}_{\Sigma}$, at which point the pressure of the surface of the collapsing star becomes infinitely large, 
whereby a
space-time singularity with finite thickness is developed. } \label{fig2}
\end{figure}

\begin{figure}[tbp]
\centering
\includegraphics[width=8cm]{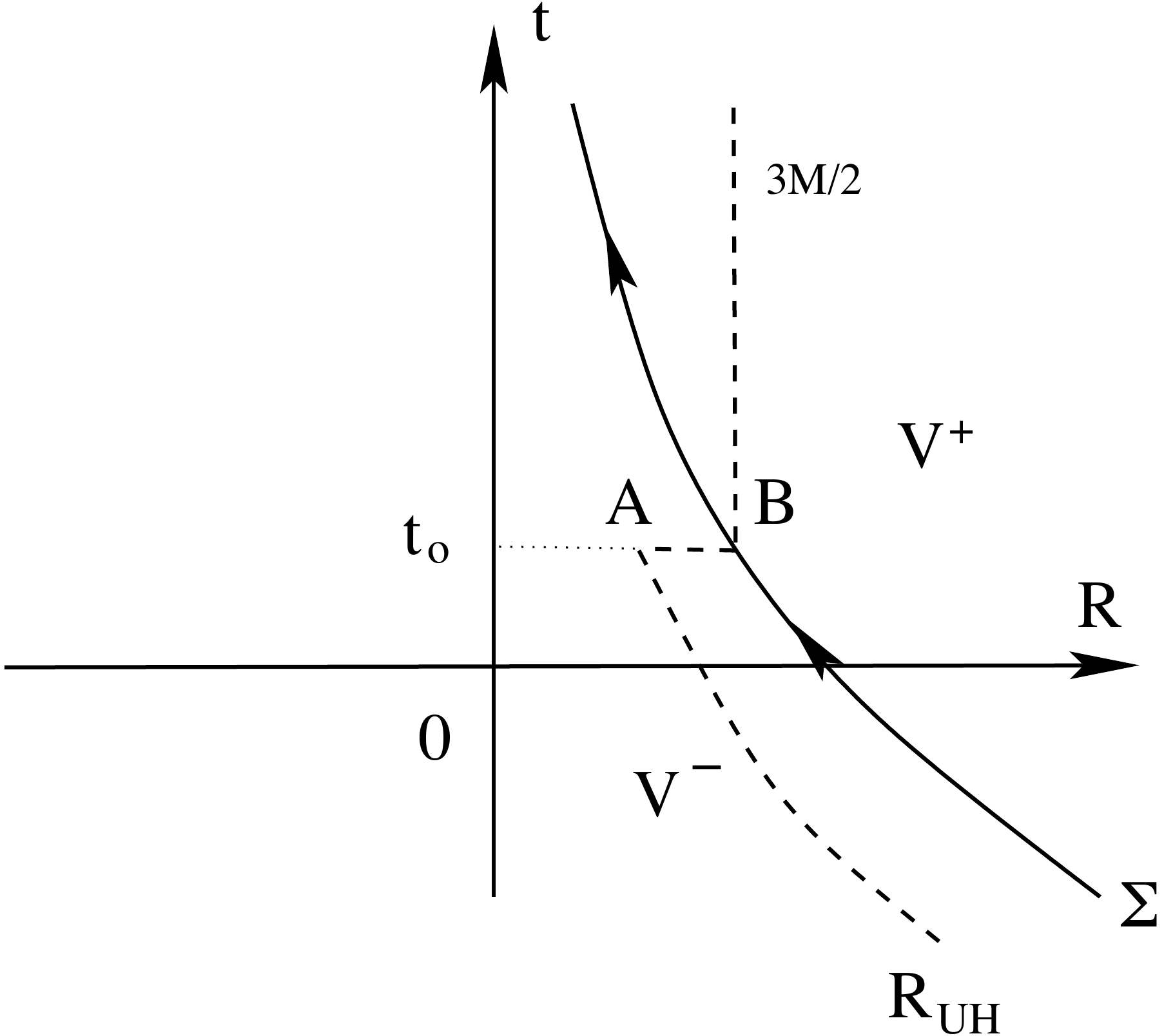}
\caption{Gravitational collapse of a spherically symmetric star with its radius $R_{\Sigma}(t)$, which divides the space-time into two regions,  $V^{\pm}$.
 The curved line $R_{UH}$ denotes the universal horizon formed inside the star, while the vertical straight line $R = 3M/2$ is the  universal horizon of the Schwarzschild
 vacuum solution. When the star collapses inside the Schwarzschild universal horizon $R = 3M/2$,  the  universal horizon suddenly jumps from $R_{UH} (t_o)$ to  $3M/2$,
 because of the non-zero mass of the collapsing surface of the star.
 } \label{fig3}
\end{figure}

\begin{figure}[tbp]
\centering
\includegraphics[width=8cm]{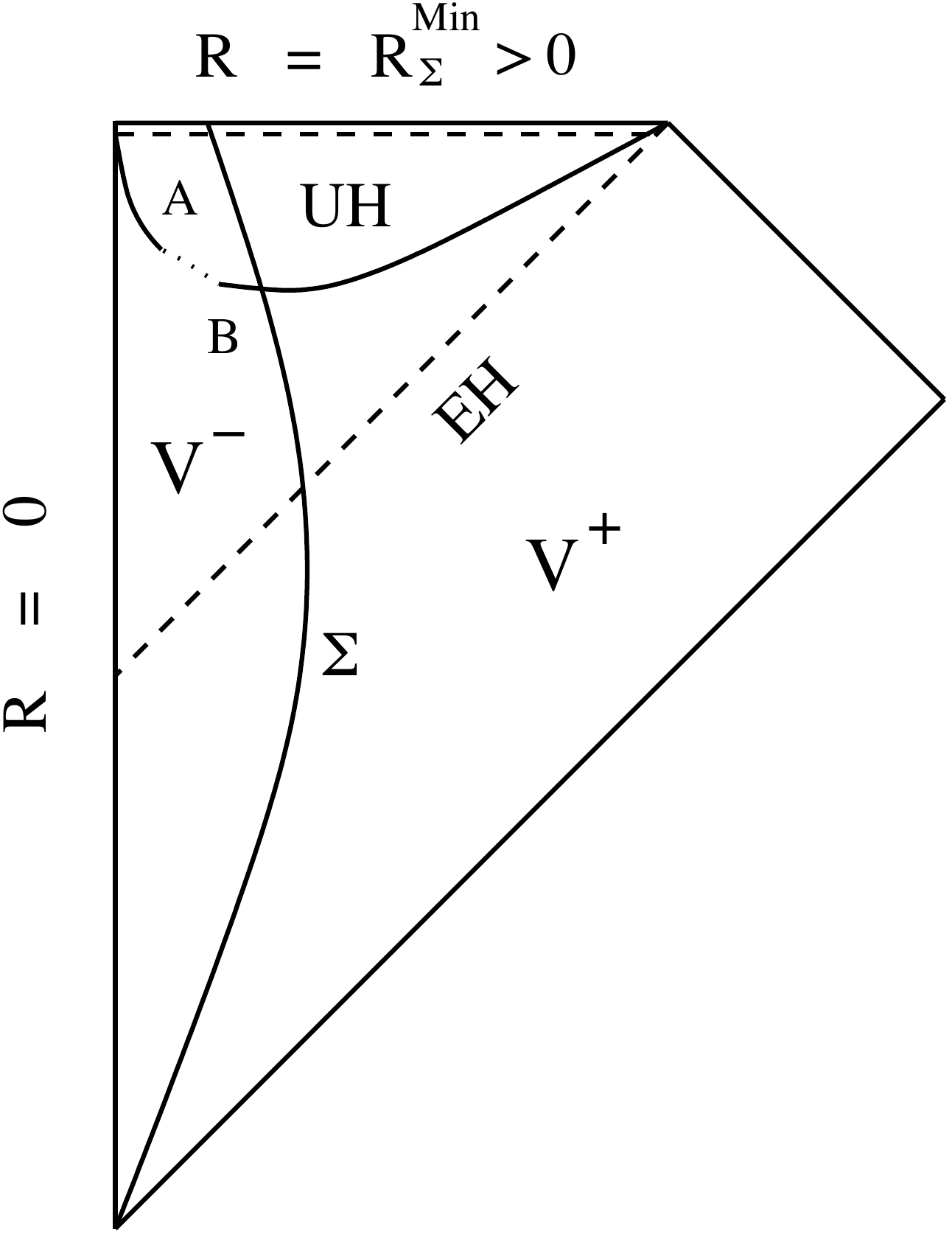}
\caption{Penrose diagram of a collapsing star described in the content  for the case where  the free parameters of the solution are chosen as $r_\Sigma=1, M = 1, a_0 = 3, 
V_0 = 0.6, H = 1.5$.  The surface $\Sigma$ of the collapsing star divides the whole space-time into two regions, the external region $V^{+}$ and the internal  region $V^{-}$. 
The curved line $UH$ denotes the universal horizon, while the straight line $EH$ the event horizon.  The star collapses to form generically a space-time singularity  at a 
finite non-zero radius $ R = R^{Min}_{\Sigma} > 0$.   } \label{fig4}
\end{figure}

It can be shown that the weak energy condition  $\sigma\geq 0,\; \sigma+\eta\geq 0$ can always be satisfied by properly choosing the free parameters involved in the solution, before the 
formation of the universal horizon. In particular, from Eq.(\ref{4.23}) we evan see that $\sigma$ is always non-negative, and
\bqn
\lb{4.23c}
\sigma + \eta =\frac{1}{8\pi R_\Sigma {\cal{G}}}\left[{\cal{G}} - \left(\frac{3M}{R_\Sigma} -1\right)\right].
\eqn
Thus, for $R_\Sigma \ge 3M$, the weak energy condition is always satisfied. When $R_\Sigma < 3M$, it is also satisfied, provided that ${\cal{G}} \ge  {3M}/{R_\Sigma} -1$, or equivalently
\bq
\lb{4.23d}
H^2R_\Sigma^2 + \frac{4M}{R_\Sigma} \ge \frac{9M^2}{R_\Sigma^2}.
\eq
Clearly,  by properly choosing the free parameters involved in the  solution,  this condition can  hold  until the moment where the whole collapsing star is inside
the universal horizon.  However, at the end     $ \tau = \tau_{Min}$ of the collapse this condition is necessarily violated, as can be
seen from Eqs.(\ref{4.23b}) and (\ref{4.23d}). Using the geometric radius   $R=r a(t)=r a_0 e^{-H t}$ inside the collapsing star,  we find that the apparent horizon 
given by Eq.(\ref{4.4}) and the universal horizon 
given by Eq.(\ref{4.12}) can be expressed
as,
\bqn
\lb{4.24}
R_{UH}&=&r_{UH} a(t)            \nb\\
&=& a_0 e^{-H t}\sqrt{\frac{2}{\sqrt{V_0^4+4V_0^2 a_0^2 H^2 e^{-2H t}}-V_0^2}}, \nb\\
R_{AH}&=&r_{AH} a(t)= \frac{1}{H}.
\eqn
In  Fig. \ref{fig1} we show one of the cases, in which the free parameters are chosen as $r_\Sigma=1, M = 1, a_0 = 3, V_0 = 0.6, H = 1.5$.
Then, we find that the weak energy condition holds until the moment $t=0.624554$, at which we  have $ R_{UH}= 1.24387 > R_\Sigma=1.1756$. That is, the weak energy condition holds
all the way down to the moment when  the whole star collapses inside the universal horizon, as it is illustrated clearly in  Fig.\ref{fig2}. In this figure, three horizontal lines, $ R = 2M, \;
3M/2, \; R^{Min}_{\Sigma}$ are also plotted, where  $R_{UH}^{\text{Sch.}} = 3M/2$   is the universal horizon in the Schwarzschild space-time [cf. Eq.(\ref{0.4})]. For the current choice of the free parameters, the 
universal horizon is not continuous. In fact, when the star collapses to the moment $t = t_o$, where $t_o$ is given by $R_{\Sigma}(t_o) = 3M/2$, the universal horizon jumps from 
$R_{UH}(t_o)$ to $3M/2$, as shown more clearly in Fig. \ref{fig3}. Physically, this is because that the surface shell of the collapsing star has non-zero mass.
The collapse ends at $t = t^{Min}_{\Sigma}$, where $R_{\Sigma}( t^{Min}_{\Sigma}) = R^{Min}_{\Sigma}$, 
at which the surface pressure  of the collapsing star becomes
infinitely large. It is remarkable that $R^{Min}_{\Sigma}$ generically is different from zero, that is, the collapse generically forms a space-time singularity that has finite radius. The corresponding Penrose 
diagram is given   in Fig. \ref{fig4}. In this figure the location of the event horizon denoted by the  straight line $EH$  is also marked, although it can be penetrated by particles with sufficiently large velocities,
and propagate to infinity, even they are initially trapped inside it. However, this is no longer the case when across the universal horizon.  As explained above, once they are trapped  inside the universal horizon,
they cannot penetrate it and propagate to infinities,  even they are moving with infinitely large velocities. As a result, an absolutely black region is (classically)  formed from the gravitational  collapse of a massive star, 
and this region  is  black,  even in theories that allow instantaneous  propagations!

\section{Conclusions}

In this paper, we have first generalized the definition of a stationary  universal horizon to a dynamical one, by simply  replacing Killing horizons by apparent ones. Then, 
we have constructed an analytical model that represents the gravitational collapse of a spherical symmetric star with finite thickness, and  shown explicitly that  dynamical universal horizons
can be formed from such a ``realistic" gravitational collapse. Here ``realistic" is referred to as a gravitational collapse of a star with a finite thickness  that satisfies at least the weak energy condition \cite{HE73}. 

To have the problem tractable, we have assumed that the star consists of an anisotropic and homogeneous perfect fluid and that outside the star the space-time is vacuum in the framework of the 
Einstein-aether theory \cite{EA}. When the effects of the aether field is negligible, the vacuum space-time is uniquely  described by the
Schwarzschild solution \cite{EA}. Even in this case, solving the khronon equation (\ref{1.11}) inside the star  is
still very complicated. Instead, we have further assumed that the velocity of the the khronon is infinitely large, so the sound horizon of the khronon coincides with the universal horizon.
Then, we have  found that an analytical solution exists for the star made of 
the de Sitter universe, and shown explicitly how a dynamical universal horizon can be formed, as can be seen from Figs. \ref{fig2} - \ref{fig4}.  

Although such a constructed  model serves our current purpose  very well,
that is, to show that universal horizons can be indeed formed from realistic gravitational collapse,  it would be very interesting to 
consider cases without (some of) the above assumptions, specially the case  in which  the space-time outside of the star is not vacuum, so that the star 
may radiate, when it is collapsing.

In addition, although in this paper we have considered the formation of the universal horizons only  in the framework of the Einstein-aether theory, it is expected that our main conclusions 
should be true  in  other theories of gravity with  broken LI, including the HL gravity \cite{ZWWS,BPS}.

\section*{Acknowledgements}

This work was done partly when A.W. was visiting the State University of Rio de Janeiro (UERJ). He would like to thank UERJ for hospitality.
It was done also during the visit of M.T. to Baylor University. M.T. would like to express his gratitude to Baylor. This work is supported in part by  
Ci\^encia Sem Fronteiras, Grant No. A045/2013 CAPES, Brazil (A.W.); NSFC Grant No. 11375153, China (A.W.);   CAPES, CNPq and FAPERJ Brazil (M.F.A.S.);
 CSC Grant No. 201208625022, 
China (M.T.);  and Baylor Graduate Scholarship (X.W.).



\begin{thebibliography}{nbound}

\bibitem{Liberati13}  A. Kostelecky and N. Russell, Rev. Mod. Phys. {\bf 83} 11 (2011) [arXiv:0801.0287v7, January 2014 Edition].

\bibitem{LZbreaking}  D. Mattingly,   Living Rev.  Relativity, {\bf 8}, 5 (2005);
S. Liberati,   Class. Qnatum Grav. {\bf 30}, 133001 (2013).

\bibitem{LACW}   K. Lin, E. Abdalla, R.-G.  Cai, and A. Wang, Inter. J. Mod. Phys. D{\bf 23},   1443004   (2014);
K. Lin, F.-W. Shu, A. Wang, and Q. Wu, Phys. Rev. D{\em in press} (2015) [arXiv:1404.3413].


\bibitem{QGs}  S. Carlip, {\em Quantum Gravity in 2+1 Dimensions},  Cambridge Monographs on Mathematical Physics (Cambridge University Press,
     Cambridge, 2003); C. Kiefer, {\em Quantum Gravity} (Oxford Science Publications, Oxford University Press, 2007).

\bibitem{Horava} P.  Ho\v{r}ava,   J. High Energy Phys. {\bf 0903}, 020 (2009);
         Phys. Rev. D{\bf 79}, 084008 (2009).

\bibitem{Lifshitz} E.M. Lifshitz,   Zh. Eksp. Toer. Fiz. {\bf 11}, 255 (1941);
       {\em ibid.}, {\bf 11}, 269 (1941).

\bibitem{Visser} M. Visser,  Phys. Rev. D{\bf 80}, 025011 (2009);
                 {\em Power-counting renormalizability of generalized  Ho\v{r}ava gravity}, arXiv:0912.4757.

\bibitem{Stelle}  K.S. Stelle,  Phys. Rev. D{\bf 16},   953 (1977).

\bibitem{Collin04}  J. Collins, A. Perez, D. Sudarsky, L. Urrutia, and H. Vucetich,
Phys. Rev. Lett. {\bf 93},  191301 (2004).

\bibitem{PT14} M.  Pospelov and C. Tamarit,   J. High Energy Phys. {\bf 01} (2014) 048.

\bibitem{ZWWS}  T.  Zhu,  Q. Wu, A. Wang, and F.-W.  Shu,    Phys. Rev. D{\bf  84}, 101502(R) (2011); T.  Zhu, F.-W.  Shu, Q. Wu, and A. Wang,
                     Phys. Rev. D{\bf 85}, 044053 (2012).

\bibitem{LMWZ}  K. Lin, S. Mukohyama, A. Wang, and T. Zhu, Phys. Rev. D{\bf 89}, 084022 (2014).

\bibitem {Will}   C.M. Will, {\em The Confrontation between General Relativity and Experiment},  Living Rev. Relativity, {\bf 9}, 3 (2006)
[http://www.livingreviews.org/lrr-2006-3; arXiv:gr-qc/0510072].

\bibitem{ZHW} T.  Zhu,   Y.-Q. Huang, and A. Wang,    Phys. Rev. D{\bf 87}, 084041 (2013);  A. Wang, Q. Wu, W. Zhao,  and T. Zhu,
     Phys. Rev. D{\bf 87}, 103512 (2013); T. Zhu, W. Zhao, Y.-Q. Huang, A. Wang, and Q. Wu,
              Phys. Rev. D{\bf 88}, 063508 (2013).

\bibitem {cosmo} E. Komatsu, {\em et al.},     Astrophys. J. Suppl. {\bf 192}, 18 (2011); P. Ade et al.,  Planck Collaboration,
Astron. $\&$ Astrophys.,  (2014).

\bibitem{JK}     S. Janiszewski and A. Karch,  JHEP, {\bf 02}, 123 (2013);  Phys. Rev. Lett. {\bf  110}, 081601 (2013).


 \bibitem{BPS} D. Blas, O. Pujolas, and S.  Sibiryakov,  Phys. Lett. B{\bf 688}, 350 (2010);  J. High Energy Phys. {\bf 1104}, 018 (2011).

\bibitem{EA} T. Jacobson and Mattingly, Phys. Rev. D{\bf 64}, 024028 (2001); T. Jacobson, Proc. Sci. QG-PH, {\bf 020} (2007).

  \bibitem{Jacob} T. Jacobson, Phys. Rev. D{\bf 81}, 101502 (R) (2010).

\bibitem{Wang13} A. Wang, {\em On ``No-go theorem for slowly rotating black holes in Ho\v{r}ava-Lifshitz gravity}, arXiv:1212.1040.


\bibitem {Yagi}  K. Yagi, D. Blas, N. Yunes, and E. Barausse,
 Phys. Rev. Lett. {\bf 112}, 161101 (2014);
 K. Yagi, D. Blas, E. Barausse, and N. Yunes,
Phys. Rev. D{\bf 89}, 084067 (2014).

\bibitem{NM} R. Narayan and J.E. MacClintock, Mon. Not. R. Astron. Soc., {\bf 419}, L69 (2012).


 
\bibitem{BS11}  D. Blas and S. Sibiryakov, Phys. Rev. D{\bf 84}, 124043 (2011).

\bibitem{UHs}  E. Barausse, T. Jacobson, and T. Sotiriou, Phys. Rev. D{\bf 83}, 124043 (2011);
 B. Cropp, S. Liberati, and M. Visser, Class. Quantum Grav. {\bf 30}, 125001 (2013);
  M. Saravani, N. Afshordi, and R.B. Mann, Phys. Rev. D{\bf 89}, 084029 (2014);
S. Janiszewski, A. Karch, B. Robinson, and D. Sommer, JHEP {\bf 04}, 163 (2014);
C. Eling and Y. Oz, JHEP, {\bf 11}, 067 (2014);
T. Sotiriou, I. Vega, and D. Vernieri, Phys. Rev. D{\bf 90}, 044046 (2014);
J. Bhattacharyya  and D. Mattingly, 	``{\em Universal horizons in maximally symmetric spaces}," arXiv:1408.6479. 



\bibitem{BBM} P. Berglund, J. Bhattacharyya, and D. Mattingly, Phys. Rev. D{\bf 85}, 124019 (2012); Phys. Rev. Lett. {\bf 110}, 071301 (2013).

\bibitem{LGSW}  K. Lin, O. Goldoni, M.F. da Silva, and A. Wang,  Phys. Rev. D {\em in press} (2015)  [arXiv:1410.6678].

\bibitem{CLMV} B. Cropp, S. Liberati, A. Mohd, and M. Visser, Phys. Rev. D{\bf 89}, 064061 (2014).

\bibitem{HE73} S.W. Hawking, G.F.R. Ellis, {\em The Large Scale Structure of Spacetime} (Cambridge University Press, Cambridge, 1973).

\bibitem{Hayd}    S.A. Hayward,  Phys. Rev. {\bf D49}, 6467 (1994);   Class. Quantum Grav. {\bf 17}, 1749 (2000).

\bibitem{Wang}  A. Wang,       Phys. Rev. D{\bf 68},  064006 (2003);  
 Phys. Rev. D{\bf 72}, 108501 (2005);  Gen. Relativ. Grav.   {\bf 37}, 1919  (2005);
 Y. Wu, M.F.A. da Silva, N.O. Santos, and A. Wang,
 Phys. Rev. D{\bf 68}, 084012 (2003);
    A.Y. Miguelote, N.A. Tomimura, and  A. Wang,    Gen. Relativ. Grav.   {\bf 36}, 1883 (2004);
  P. Sharma, A. Tziolas, A. Wang, and Z.-C. Wu,       Inter. J. Mord. Phys. A{\bf 26}, 273 (2011).

\bibitem{IC66} W. Israel, Nuovo Cimento, B 44 (1966) 1; B 48 (1967) 463 (E); A. Wang and N.O. Santos, Inter. J. Mod. Phys. A{\bf 25}, 1661 (2010).


\bibitem{CS97}  W.B. Bonnor, A.K.G. de Oliveira, and N.O. Santos, Phys. Rept. {\bf 181}, 269 (1989).  

\bibitem{Cosmo} L. Amendola  and S. Tsujikawa, {\em Dark Energy: Theory and Observations} (Cambridge University Press,  Cambridge, 2010).




\end{thebibliography}
\end{document}